# Choice modelling in the age of machine learning

Discussion paper

Sander van Cranenburgh[I*], Shenhao Wang[II], Akshay Vij[III], Francisco Pereira[IV], Joan Walker[V]

[I] *Transportation and Logistics group, Department of Engineering Systems and Services, Delft University of Technology*
[II] *Urban Mobility Lab, Massachusetts Institute of Technology*
[III] *Institute for Choice, University of South Australia*
[IV] *Department of Technology, Management and Economics, Technical University of Denmark*
[V] *Department of Civil and Environmental Engineering, University of California Berkeley*
[*] *Corresponding author. s.vancranenburgh@tudelft.nl*

*Abstract*

*Since its inception, the choice modelling field has been dominated by theory-driven modelling approaches. Machine learning offers an alternative data-driven approach for modelling choice behaviour and is increasingly drawing interest in our field. Cross-pollination of machine learning models, techniques and practices could help overcome problems and limitations encountered in the current theory-driven modelling paradigm, such as subjective labour-intensive search processes for model selection, the inability to work with text and image data. However, despite the potential benefits of using the advances of machine learning to improve choice modelling practices, the choice modelling field has been hesitant to embrace machine learning. This discussion paper aims to consolidate knowledge on the use of machine learning models, techniques and practices for choice modelling, and discuss their potential. Thereby, we hope not only to make the case that further integration of machine learning in choice modelling is beneficial, but also to further facilitate it. To this end, we clarify the similarities and differences between the two modelling paradigms; we review the use of machine learning for choice modelling; and we explore areas of opportunities for embracing machine learning models and techniques to improve our practices. To conclude this discussion paper, we put forward a set of research questions which must be addressed to better understand if and how machine learning can benefit choice modelling.*

## 1 Introduction

The development of the Random Utility Maximisation (RUM) model (McFadden 1974) in the mid-1970s has been foundational for the way in which choice behaviour has been modelled and studied over the past 50 years (Hess and Daly 2014). To develop a statistical model of choice behaviour, in this theory-driven modelling paradigm the analyst imposes structure on the data by postulating that decision makers make decisions based on utility theory or some variation thereof. Recently, an alternative modelling paradigm is gaining ground in the choice modelling field. In this data-driven modelling paradigm – also referred to as machine learning – the structure of the problem is learned from the data, as opposed to being imposed by the analyst based on prior beliefs or behavioural theories. A growing body of literature is emerging with studies that bring models, estimation techniques and practices from machine learning to the choice modelling field (Wong et al. 2017; Sifringer et al. 2020; Lederrey et al. 2021).

The motivations put forward in these studies to employ machine learning for choice modelling are diverse but may be summarised in terms of four main points. Firstly, machine learning models can overcome problems of theory-driven choice models relating to the search for the optimal model specification, and the adverse effects caused by model misspecification. In theory-driven choice models, the analyst imposes structure to the problem through functional forms and variable selections, based on theoretical frameworks. Then, in a time consuming and often ad hoc and subjective process, the final model is selected from a series of competing specifications (Paz et al. 2019; Rodrigues et al. 2020). In case this final model turns out to be a poor descriptor of the true underlying data-generating process, model inferences and predictions can be erroneous. In contrast, machine learning models learn the structure of the problem from the data, without any prior theoretical assumptions about the data generating process. This can make the process of model selection more efficient and less susceptible to subjective biases. Moreover, it has been argued that learning the structure from the data has the additional advantage of offering the possibility to find the unexpected. Secondly, machine learning models often achieve higher goodness-of-fit than their theory-driven counterparts, especially in the context of prediction applications (Lee et al. 2018). While goodness-of-fit is seldom an aim in and of itself in the choice modelling field, better fit is generally desirable as it is taken as a signal that the model has accurately captured the underlying choice process. Thirdly, machine learning models and estimation techniques often work comparatively well in combination with large and continuous streams of data (Danaf et al. 2019). As choice modellers increasingly get access to very large data sets, machine learning offers opportunities to develop new ways for mining behaviourally meaningful, statistically robust and computationally efficient insights from these data sets. Finally, machine learning models can work with types of data that are currently outside the realm of traditional theory-driven discrete choice models, such as text and image data. Introducing machine learning to choice modelling thus opens up the opportunity to extend the reach of choice modelling to model choice behaviour using different data types (Van Cranenburgh 2020). In summary, there is clear potential for machine learning models, estimation techniques and practices to enrich choice modelling practices.

However, despite this potential the choice modelling field has been somewhat hesitant to embrace machine learning. This hesitance appears to be attributable to at least three related factors. Firstly, there seems to be a lack of understanding about machine learning, for instance, about what theory-driven choice models and machine learning have in common and what sets them apart. Possibly, this lack of understanding is caused by differences in vocabulary and terminology across the two fields (Breiman 2001), which impedes choice modellers from effectively understanding the machine learning literature. Secondly, there exist persistent misconceptions about machine learning among choice modellers that are potentially holding back analysts from being open to what machine learning could offer. Common misconceptions are, for instance, that machine learning models can only be used for prediction as opposed to behavioural inference, and that machine learning models are overfitting the data more often than not. Thirdly, which is in part a consequence of the above two factors, there seems to be a lack of recognition of the potential value of integrating machine learning models, techniques and practices for the choice modelling field.

This discussion paper aims to consolidate knowledge on the use of machine learning models, techniques and practices for choice modelling, and discuss their potential for improving current practices. With this discussion paper we hope (1) to convince choice modellers that further integration of machine learning in choice modelling is beneficial; and (2) to facilitate (further) integration.

The remainder of the paper is structured as follows. Section 2 starts by clarifying the similarities and differences between the two modelling paradigms, and sets the stage by providing a concise literature overview of existing applications of machine learning in the choice modelling field. To understand where machine learning could or could not impact choice modelling practices, Section 3 reinforces the strengths of the current theory-driven choice models. In this section, we specifically ask ourselves 'what makes the current theory-driven modelling paradigm strong?' and 'how does it compare to the machine learning paradigm?' Having identified the strengths of the current theory-driven paradigm, Section 4 identifies key areas of opportunity where choice modelling could benefit from embracing machine learning. Finally, Section 5 concludes with a discussion on the road ahead. It delves into the bigger research questions that must be addressed to assess if and how machine learning could transform choice modelling practices.

## 2 Similarities and differences between theory-driven and data-driven modelling

Inspired by the seminal paper on data-driven versus theory-driven models of Breiman (2001), several papers have tried to position data-driven models with respect to other adjacent fields. In this tradition, this section aims to explicate the similarities and differences between theory-driven choice models and data-driven (machine learning) models, while specifically taking a choice modeller's perspective. We focus on high-level similarities and differences between the approaches that have motivated the development of models in either discipline.

The realm of machine learning models is so broad that it hinders making general statements. Therefore, we first narrow down what we mean by machine learning models in this paper. When we talk about machine learning models, we specifically have in mind machine learning classifiers that are commonly encountered in the computational intelligence field. These include, among others, Artificial Neural Networks (ANNs), Boltzmann Machines (BMs), Support Vector Machines (SVMs), Bayesian Networks (BNs), Probabilistic Graphical Models (PGMs), Decision Trees (DT), Association Rules (AR) and ensemble methods, such as Random Forests (RF) and Gradient Boosting (GB). Likewise, when we talk about theory-driven discrete choice models, we specifically have in mind choice models that are underpinned by behavioural theories, such as Utility Theory (UT) (McFadden 2001), Regret Theory (RT) (Loomes and Sugden 1982), and Prospect Theory (PT) (Kahneman and Tversky 1979; Tversky and Kahneman 1992). These models are almost without exception estimated in logit, probit, mixed logit and latent class forms.

Furthermore, in the discussions that follow, we mostly focus on comparisons between theory-driven choice models and *supervised* machine learning. Supervised machine learning is one of the three main subdomains of machine learning: the other two being unsupervised learning and reinforcement learning. Supervised learning is concerned with learning a function that maps input features (i.e., the explanatory variables) to an output (i.e., the dependent variable) (Murphy 2012). Hence, in supervised learning, the dependent variable is part of the data. Common supervised learning tasks are classification and regression tasks. As choices can be cast as mutually exclusive classes, choice modelling can be seen as a classification task. Unlike supervised learning, unsupervised learning is concerned with drawing inferences from data that does not have an independent variable (Murphy 2012). One of the most common unsupervised learning methods is cluster analysis, which is used to find (hidden) patterns or groupings in data. Furthermore, as a variation on the theme, there is also semi-supervised learning. Semi-supervised learning conceptually sits between supervised and unsupervised learning and operates on partially labelled data. It is commonly used when unlabelled data are widely available, but labels are expensive to obtain (van Engelen and Hoos 2020). Finally, reinforcement learning is concerned with how agents can learn to take actions in an environment in order to maximize the notion of cumulative delayed reward (Jo 2021). For instance, when training a computer to play chess the reward only comes after the game is won. Readers interested in a deeper discussion on the various learning techniques are referred to Jo (2021).

## 2.1 Similarities

Theory-driven choice modelling and supervised machine learning have much in common. Both are grounded in statistical theory. Researchers in both fields aim to develop models to generate predictions and inferences that are interpretable, replicable, scalable, flexible (allowing for necessary complexity), and robust. Unsurprisingly, in both areas, concepts like random variable, probability distribution, error term, confidence and prediction intervals, estimator consistency and asymptotic properties, central limit theorem, latent variable, etc. are cornerstones concepts. While many such concepts are used in both fields, they often come under different names. To make the machine learning literature more accessible to choice modellers, Table 1 provides an overview of the most common shared concepts.

Likewise, it is not surprising to see that both fields face some of the same challenges, relating to e.g. endogeneity, explainability, heterogeneity, unbalanced datasets, and the trade-off between model complexity and generalizability. Facing some of the same challenges is not to say that each challenge is equally important to both fields, or that it is discussed and dealt with in the same manner. For instance, there are few if any explicit references to 'endogeneity' in the machine learning literature. Nonetheless, over the last couple of years there has been a strong uptake in papers focussing on aspects that relate to, or are affected by, endogeneity, such as on fairness and bias issues and correlated error terms with input variables (Mehrabi et al. 2019). Conversely, explainability has hardly been mentioned as a challenge in choice modelling, while it has gathered significant attention in machine learning in recent years. Yet, it goes without saying that explaining and understanding the model and its outcomes is essential to both fields.

**Table 1: Shared concepts**

| Terminology in … | |
|---|---|
| ***Choice modelling*** | ***Machine learning*** |
| *Alternative* | *Output class* |
| *ASC* | *Intercept* |
| *Attribute, covariate* | *Feature, input* |
| *Binary Logit function* | *Sigmoid function* |
| *Efficient experimental design* | *Active learning* |
| *Estimation* | *Training* |
| *Full information maximum likelihood estimation* | *Batch gradient descent training* |
| *Hit rate* | *Accuracy* |
| *Log-likelihood* | *Log loss* |
| *Model parameter* | *Weight* |
| *Multinomial Logit function* | *Softmax function* |
| *Observation* | *Example, instance* |

## 2.2 Differences

What differentiates machine learning from choice modelling is, arguably, more interesting. A sharp distinction cannot be drawn, in part because clear definitions for what constitute a theory-driven approach and what constitute a data-driven approach are missing (Erdem et al. 2005). Nonetheless, there is a fundamental epistemological difference between the two fields, with a wide range of implications.

For a theory-driven (choice) modeller, the guiding principle is that based on theory choice models can be specified that enable making behavioural inferences and predictions. The theory postulates how a set of the explanatory variables, $X$, relate to the choice, $Y$ (Reiss and Wolak 2007). Essentially, the theory provides the first principles that describe the choice behaviour in a parsimonious way, in much the same way as e.g. the laws of classical mechanics provide the first principles of the motion of macroscopic objects (Ran and Hu 2017). For a theory-driven choice modeller, competing models originate from different decision theories, like Utility Theory (UT), Prospect Theory (PT), Regret Theory (RT), etc. and different functional forms, e.g., different error term distributions. After having identified the best model from a set of competing models, the analyst conceives this model as the *true* representation of the data from which behavioural inferences can be made. After all, the parameters of a theory-driven discrete choice model are only meaningful under the assumption that the model, and the theory it is built on, are correct. For instance, after having estimated a linear-additive RUM model, the analyst interprets the parameters as marginal utilities (Ben-Akiva and Lerman 1985). Modellers using theory-driven discrete choice models mostly focus on formalising understanding of how decision makers make choices, e.g. whether the average decision maker is loss averse or not. The theory-driven discrete choice models they use are also understood as causal models. The direction of causation (i.e. the underlying mechanism giving rise to observed statistical dependencies) is derived from the behavioural theory. As a result of their causal structure, they are generally deemed suitable to make predictions – e.g. for policy interventions – beyond the support of the observed data, or for counterfactual analysis (Reiss and Wolak 2007).

For a machine learning analyst, the guiding principle is that the data generating process is complex, mysterious and at least partly unknowable (Breiman 2001); or, as Ran and Hu (2017) put it "the underlying 'first principles' are unknown, or the systems under study are too complex to be mathematically described". Therefore, a machine learning analyst puts the data at the centre and is not concerned with understanding the first principles, or 'true' underlying mechanisms that generate the data. For a machine learning analyst, the model is as truthful to a phenomenon as the model is capable of generalising to out-of-sample data. As a machine learning analyst typically builds a model with the aim of identifying the best course of action (e.g. best recommendation), they concentrate on prediction (Bzdok et al. 2018). The predictive accuracy is the capability of the learned model to provide accurate predictions for further data which are coming from the same data generating process (Murphy 2012). As a result of the strong focus on prediction, machine learning models tend to be highly flexible – in the sense that they have many parameters and make few a priori assumptions about the data generating process. This flexibility gives machine learning models the capacity to learn the structure of the data while not being constrained by restrictive assumptions. Commonly, the only real assumption made is that the data on which the model is trained are drawn i.i.d. from an unknown multivariate distribution (Breiman 2001). Furthermore, again due to the strong focus on prediction, machine learning models usually only learn the statistical associations between the variables; they do not seek to learn causal structural relationships (Schölkopf et al. 2021). Understanding of the underlying structural mechanisms that generate the data is considered to be of secondary importance to most machine learning analysts (Bzdok et al. 2018).

As a consequence of these disparate guiding principles, choice modelling and machine learning communities have given different weights to different concepts, have focussed on different types of models, and have developed different practices. For example, the concept of model interpretability has always been fundamental to theory-driven choice models, whereas it has mostly been more of a second thought in machine learning given its focus on prediction as opposed to inference. Only recently, model interpretability has started to receive attention. Likewise, the concept of model identifiability is paramount in both choice modelling and machine learning, but for different reasons. Identifiability of a model means that there are no two different sets of estimable parameters that give the same probability distribution function on any data, or in other words are observationally equivalent (Rothenberg 1971). In theory-driven discrete choice models, since the theory provides meaning to the model parameters, models that are not uniquely identifiable are not unambiguously interpretable. Moreover, identifiability is a prerequisite for statistical inference. Lack of identifiability precludes calculation of standard errors for the parameter estimates and limits the ability to perform formal statistical hypothesis tests. Yet, statistical tests play an essential role during the model building (and inference) phase in theory-driven choice modelling. In other words, lack of identifiability undermines the principle that the true data generating process can be inferred by testing statistical models. Therefore, models that are not uniquely identifiable are to be avoided by theory-driven discrete choice modellers. Identifiability (or rather the lack thereof) is also important for machine learning, as the lack of identifiability complicates developing learning and algorithm theories. (see Ran and Hu 2017 for a comprehensive discussion on this topic). For instance, it hinders analysing properties of estimates and poses severe challenges when it comes to statistical testing of competing models (Horel and Giesecke 2019). Classic asymptotic statistics, like the Likelihood Ratio Statistic (LRS), Akaike Information Criterion (AIC) and Bayesian Information Criterion (BIC) (Schwarz 1978) are not (widely) used in machine learning for this reason.

Choice modelling and machine learning have developed different software and data practices too. Modelling practices in choice modelling are typified by their transparency. Choice models usually consist of a few well-known building blocks, e.g., a logit kernel, a membership function, a mixture kernel, a random number generator, etc. These building blocks are incorporated in a few widely used packages, such as (Pandas) Biogeme (Bierlaire 2018) and Pylogit (Brathwaite and Walker 2018) for Python and GMNL (Sarrias and Daziano 2017) and Apollo (Hess and Palma 2019) for R. As a result, the height of the software pyramid is fairly short, and the software is fairly transparent. Choice modellers typically rebuild each other's models from scratch, i.e. based on the equations provided in the respective paper.

Modelling practices in machine learning are typified by a much larger set of building blocks. That is, researchers build pieces of software, such as new types of layers for neural networks, which they share with their peers. Commonly used software platforms by machine learning researchers are TensorFlow (Abadi et al. 2016), scikit-learn (Pedregosa et al. 2011) and PyTorch (Paszke et al. 2017). These peers, in turn, re-use building blocks and combine them with other blocks to assemble new models. This practice leads to a higher software pyramid, in which a researcher may have limited awareness of what is happening with models at the base of the pyramid. The machine learning community accepts this opaqueness in exchange for flexibility, while imposing strict replicability (shared golden datasets, open access publishing) and validation principles, as well as peer-review code quality control (open-source code sharing). Large machine learning conferences even have 'reproducibility' programmes (Pineau et al. 2020). When machine learning researchers rebuild each other's models, it is typically done by sharing such building blocks through software repositories, like Github.

### 2.3 Machine learning in choice modelling: A brief overview of the emerging literature

To acquire a sense of the emerging literature of studies using machine learning for choice modelling, we conducted a brief literature review. Specifically, we searched the literature for the Boolean combination of the search tags 'machine learning' and 'discrete choice model', where the tag 'machine learning' was also replaced by particular types of machine learning models, such as 'decision tree', 'artificial neural network' and 'support vector machine', etc., and the search tag 'discrete choice model' was also replaced with 'MNL' and 'logit'. Furthermore, we limited our search to four main publication outlets: Journal of Choice Modelling (JOCM), Transportation Research Part C (TrC), Transportation Research Records (TRR) and Expert Systems with Applications (ESWS). We note that two of the four selected outlets are specifically dedicated to transportation. This search criterion could therefore lead to an overrepresentation of transportation research in our overview. However, judging from an unrestricted search, it is our impression that this search criterion does not skew our main results. The search task was performed using Google Scholar. After retrieving the search results, we assessed the identified studies on whether or not a machine learning model or technique is used for discrete choice analysis. We used a fairly strict interpretation of discrete choice analysis, in the sense that the choice behaviour needs to be explained by attributes of the alternatives. Thus, studies that for instance use machine learning for mode choice or travel purpose detection from GPS traces are excluded. Using this approach, a total of 28 studies are identified, see Table 2.

The resulting list is not meant to be exhaustive, rather it should be seen as the tip of the iceberg.[1] We are aware of several early studies that use machine learning models for choice modelling, such as Nijkamp et al. (1996); Hensher and Ton (2000), which did not end up in our overview because of our outlet scope. Also, there are numerous studies not in our overview because they are still in open-access archives or in conference proceedings, such as Krueger et al. (2019); Pereira (2019); Han et al. (2020); Wang et al. (2021).

In Table 2 for each study we report the year of publication, journal, application area, machine learning model(s), programming language, and the type of data (RP or SC).[2] Additionally, to gauge the methodological progress, we report the 'methodological objective' of the study. That is, we make a distinction between studies whose primary aim is to make comparisons between machine learning and theory-driven choice models in terms of model performance (such as e.g., cross entropy, rho square, and prediction accuracy) and studies that aim to go beyond comparisons of performance. The latter type of studies, for instance, try to integrate discrete choice models and machine learning models (as to get the best of both worlds), or try to extend machine learning approaches such that they become useful to tackle challenges of choice modellers (instead of machine learning researchers).

In Table 2, we also report the type of discrete choice model that is used as the benchmark. Depending on their methodological objective, studies tend to use the benchmark discrete choice model in different ways. Studies having a methodological objective to 'compare', typically use the benchmark model to show how much better the machine learning model can do in terms of model fit and prediction accuracy; studies with a methodological objective 'beyond comparison' typically use the benchmark model to build trust in the substantive outcomes of the machine learning model. In other words, they use the benchmark models to see whether or not similar substantive outcomes (such as e.g., value-of-time estimates) are obtained. The type of discrete choice model that is used as the benchmark matters, in particular for those studies that have made comparisons in terms of model performance as their methodological objective. After all, ceteris paribus, Mixed Logit model generally attains a considerably higher model fit than its MNL cousin.

Based on Table 2 we can obtain a number of insights. Firstly, most papers are very recent. That is, 21 out of the 28 papers are published in or after 2017. This signals strong uptake in machine learning studies in the choice modelling field in recent years. Secondly, more than half of the studies look at travel mode choice behaviour. It is unclear why mode choice has received so much more attention than other types of choices. Possibly this is due to the availability of a few relatively large open RP mode choice data sets, such as Hillel et al. (2018). Thirdly, in terms of the type of machine learning model, we see that most studies use ANNs (in a variety of forms), followed by SVMs and decision trees and other rule-based models. This large variety of models suggests that no consolidation has yet taken place in terms of what particular machine learning models are suited for choice modelling. Fourthly, regarding programming languages, Python is most widely used, although R and MATLAB[3] also attain considerable market shares. Python is especially popular because there are many good machine learning libraries available, such as Keras, PyTorch, Scikit-learn, Tensorflow, Theano, Weka, etc. Fifthly, we see that most studies use revealed data. About a fifth use stated choice data. This shows that stated choice data are not precluded from being analysed using machine learning models due to their generally smaller size.

---

[1] Readers interested in a more extensive literature review are referred to Hillel et al. (2021), who focus on the literature employing machine learning models specifically for modelling passenger mode choices.
[2] Note that for ease of interpretation, for each column in the table a pie charts is made, placed at the bottom of the table.

We see that slightly over half of the studies have the methodological objective 'compare', and half the studies have the methodological objective 'beyond compare'. Earlier studies are more likely to have focussed on making comparisons in terms of model performance, while more recent studies are more likely to go beyond these comparisons. This suggests the emerging sub field is making methodological progress. Most of the studies that focus on comparisons report that machine learning models outperform their discrete choice model counterparts in terms of out-of-sample fit. But, considering that machine learning models do not carry the 'burden' (i.e., restrictions) imposed by behavioural theory (and that comparisons are mostly made with MNL models, see next paragraph), this is hardly surprising. Looking more closely at the topics of the studies that 'go beyond', we observe that they are addressing a wide variety of classic choice modelling topics, ranging from model specification, estimation, inference of economic outputs to behavioural phenomena and decision rules.

Regarding benchmarks models, we see that most studies consider the MNL model as the benchmark. This is somewhat surprising. While MNL models are still considered the workhorses of choice modelling, at present the state-of-practice in choice modelling is the (panel) Mixed Logit model (Hess 2010). Using the MNL model as benchmark is especially inexpedient for studies with a comparison objective. Since the Mixed logit model usually considerably outperforms the MNL model (Shen 2009; Keane and Wasi 2013), it is currently unclear how much is really gained by using machine learning models in terms of model performance and prediction accuracy, relative to state-of-the-practice discrete choice models.

Altogether, we conclude that as a field we are probably at the early stages of the process of integration of machine learning in choice modelling. The number of papers using machine learning for choice modelling seems still to be on the rise. The fact that many studies cited in this paper are currently only available through preprint servers and repositories (see the reference list) suggests that we can expect many more papers on this topic will be published in the near future. Furthermore, it seems that consolidation has not yet started to take place. Our field has not yet notably endorsed particular types of machine learning models (e.g. ANNs, DTs, SVMs), or has established good standard methodological practices, such as benchmark machine learning models, applications, or data sets (Hillel et al. 2021). Looking more closely into the current stream of studies, we also do not yet see signs of saturation of ideas to bringing together machine learning and choice modelling. The shift in focus from comparisons towards a deeper kind of integration is thus likely to continue. Given this early stage, it makes it opportune to (re)discover the core merits of the current theory-driven modelling paradigm, and explore areas of opportunity for embracing machine learning for choice modelling. This is done in the next two sections.

---

[3] Note that strictly speaking MATLAB is not a programming language, but a software package.

**Table 2: Identified literature**

| Author(s) | Year | Journal | Application area | Machine Learning model | Programming | Data | Methodological focus | Benchmark model |
|---|---|---|---|---|---|---|---|---|
| Mohammadian, A. & Miller, E. | 2002 | TRR | Vehicle type | ANN | Neuro solutions | RP | Compare | NL |
| Xie, C., Lu, J. & Parkany, E. | 2003 | TRR | Travel mode | ANN, DT | Matlab, C5.0 | RP | Compare | MNL |
| Cantarella, G.E. & de Luca, S. | 2005 | TR-C | Travel model | ANN | Matlab | RP | Compare | MNL, NL |
| Zhang, Y. & Xie, Y. | 2008 | TRR | Travel mode | ANN, SVM | Matlab | RP | Compare | MNL |
| Tortum, A., Yayla, N. & Gökdağ, M. | 2009 | ESWA | Travel mode | ANN, ANFIS | Matlab, Statistica | RP | Compare | MR, MNL |
| Lu, Y. & Kawamura, K. | 2010 | TRR | Travel mode | CAR | Not reported | RP | Beyond compare | N/A |
| Omrani, H., Charif, O., Gerber, P., Awasthi, A. & Trigano, P. | 2013 | TRR | Travel mode | ENN, ANN, DT, KNN, SVM | R | RP | Compare | MNL |
| Hagenauer, J. & Helbich, M. | 2017 | ESWA | Travel mode | ANN, NB, SVM, DT, RF | R | RP | Compare | MNL |
| Wong, M., Farooq, B. & Bilodeau, G.-A. | 2017 | JOCM | Financial product | RBM | Python | RP | Beyond compare | N/A |
| Alwosheel, A., van Cranenburgh, S. & Chorus, C. G. | 2018 | JOCM | Travel mode | ANN | Python | SC | Beyond compare | MNL |
| Shi, H. & Yin, G. | 2018 | JOCM | Travel mode, Horse race | Boosting MNL | R | RP | Beyond compare | MNL |
| Sun, Y., Jiang, Z., Gu, J., Zhou, M., Li, Y. & Zhang, L. | 2018 | TR-C | Train ticket | ANN, SVM | Matlab | RP | Compare | MR |
| Lee, D., Derrible, S. & Pereira, F. C. | 2018 | TRR | Travel mode | ANN | Python | RP | Compare | MNL |
| Wang, F. & Ross, C. L. | 2018 | TRR | Travel mode | XGB | R | RP | Compare | MNL |
| Alwosheel, A., van Cranenburgh, S. & Chorus, C. G. | 2019 | JOCM | Travel mode | ANN | Python | RP | Beyond compare | MNL |
| Paz, A., Arteaga, C. & Cobos, C. | 2019 | JOCM | Electricity plan, Vehicle type | SA | R | SC | Beyond compare | ML |
| Lhéritier, A., Bocamazo, M., Delahaye, T. & Acuna-Agost, R. | 2019 | JOCM | Flight booking | RF | Python | RP | Compare | MNL, LC-MNL |
| Zhao, H., Meng, Q. & Wang, Y. | 2019 | TR-C | Container slot booking | ANN | Not reported | RP | Compare | MNL |
| Van Cranenburgh, S. & Alwosheel, A. | 2019 | TR-C | Decision rules | ANN | Matlab | SC | Beyond compare | LC-ML |
| Lee, D., Mulrow, J., Haboucha, C. J., Derrible, S. & Shiftan, Y. | 2019 | TRR | Vehicle type | GBM | Not reported | SC | Beyond compare | MNL |
| Wong, M. & Farooq, B. | 2020 | TR-C | Travel mode + travel distance | Bi-partitie generative model | Python | RP | Beyond compare | N/A |
| Wang, S., Mo, B. & Zhao, J. | 2020 | TR-C | Travel mode, Train type | ANN, SVM, NB, KNN, DT, QDA | Python | SC | Compare | MNL, NL, MR |
| Wang, S., Wang, Q. & Zhao, J. | 2020 | TR-C | Travel mode | ANN | Python | RP, SC | Beyond compare | MNL |
| Newman, J. & Garoow, L. | 2020 | TRR | Airline itinerary | GBM | Python | RP | Beyond compare | MNL, NL (OGEV) |
| Yao, R. & Bekhor, S. | 2020 | TR-C | Route | KNN, RF | Not reported | RP | Beyond compare | MNL |
| Zhu, Z., Sun, Y., Chen, X., Yang, H. | 2021 | TR-C | Taxi service type | BSTF | Not reported | RP | Beyond compare | ANN, DT, NB, RF, SVM |
| Lederrey, G., Lurkin, V., Hillel, T., Bierlaire, M. | 2021 | JOCM | Travel mode | N/A | Python | RP | Beyond compare | N/A |
| Wong, M., Farooq, B. | 2021 | TR-C | Travel mode | ANN | Python | RP | Beyond compare | MNL |

| Abbreviations | |
|---|---|
| ESWA | Expert Systems with Applications |
| JOCM | Journal of Choice Modelling |
| TRR | Transportation Research Record |
| TR-C | Transportation Research Part C |
| ANFIS | Adaptive Neuro-Fuzzy Inference System |
| ANN | Artificial Neural Network (shallow & deep) |
| BSTF | Bayesian Supervised learning Tensor Factorisation |
| CAR | Class Association Rules |
| DT | Decision Tree |
| ENN | Extreme Neural Network |
| GBM | Gradient boost model |
| KNN | K-means Nearest Neighbour |
| LC | Latent Class |
| ML | Mixed Logit |
| MNL | Multinomial Logit |
| MR | Multiple Regression |
| NB | Naïve Bayesian |
| NL | Nested Logit |
| OGEV | Ordered generalized extreme value |
| QDA | Quadratic Discriminatory analysis |
| RBM | Restricted Boltzmann Machine |
| RF | Random Forest |
| SA | Simulated Annealing |
| XGB | Extreme Gradient Boost |
| RP | Revealed Preference data |
| SC | Stated Choice data |

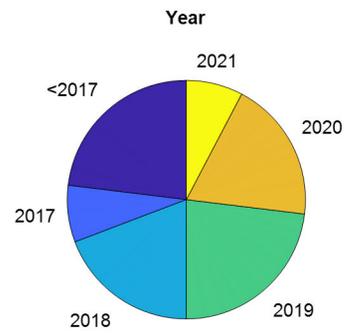
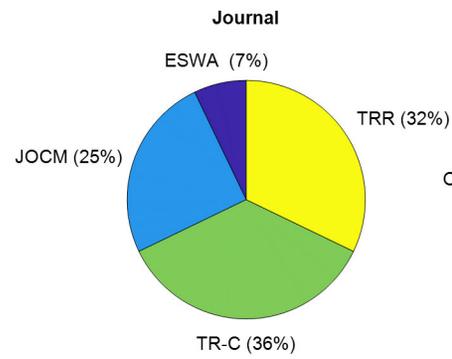
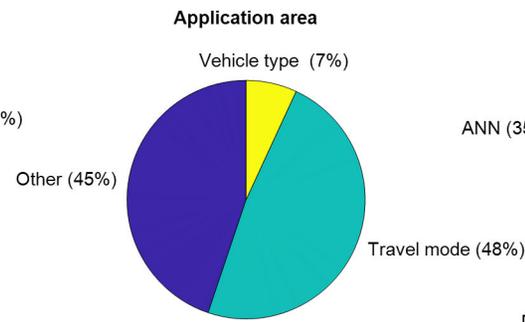
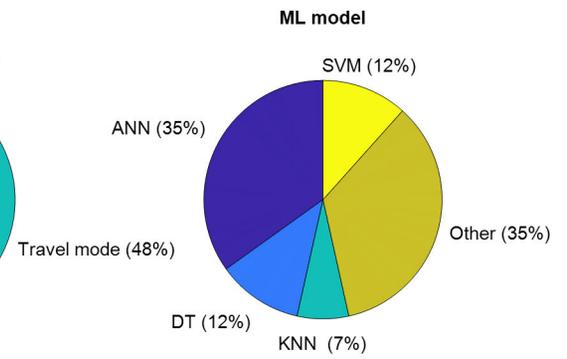
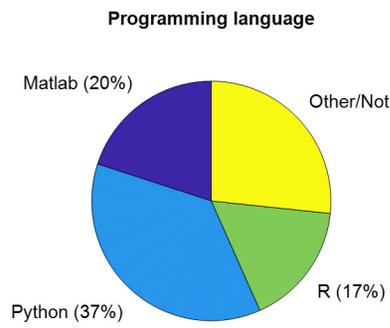
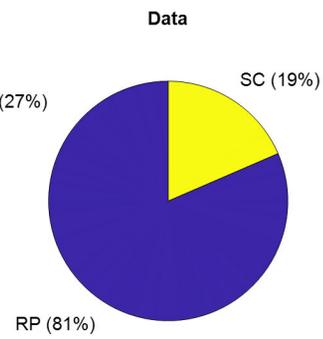
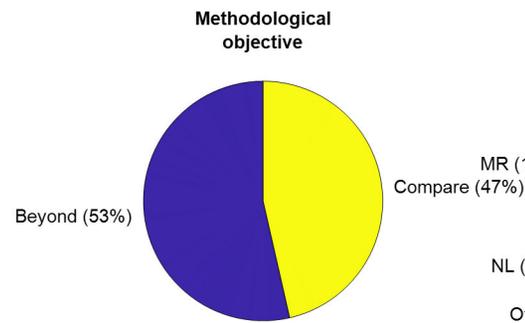
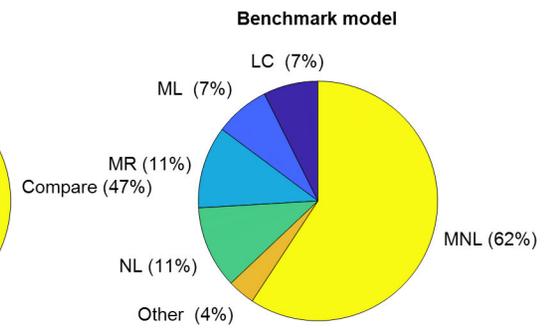

# 3 Theory-driven discrete choice models in the age of machine learning

To (re)discover the core merits of the current theory-driven modelling paradigm, in this section we ask ourselves the following questions: '*what makes the current paradigm strong?*' and '*how does it compare to machine learning?*' In answering these questions, we aim to understand where machine learning can and cannot meaningfully contribute to improving choice modelling practices.

In our view, the popularity of theory-driven discrete choice models can be ascribed to two main substantive factors: their links with behavioural theory and their close connection to stated choice experiments. Below, we discuss both factors, as well as how they compare and relate to machine learning. Finally, it is worth noting that aside from these two substantive factors that co-explain the popularity of the current theory-driven paradigm, there are likely other social factors at play as well, such as habits, inertia in the field, know-how by researchers, reputations and interests of scholars that are tied to the current modelling paradigm, existing publishing practices and outlets that are beneficial for the work within the existing paradigm, expectations of stakeholders, etc. In this section, we do not dwell upon these types of social factors, as sociology of scientific knowledge lies outside our expertise and the scope of this paper.

## 3.1 In theory we trust

The link with behavioural theory underlies the core premise of current discrete choice models, most notably the canonical RUM model – which after its inception quickly became a widely used practical tool for policy analysis and planning studies. The RUM model was the result of a series of seminal breakthroughs in choice theory and micro-econometrics (McFadden 2001). For instance, Samuelson (1948) conceptualised choice as a signal of cardinal utility coming from an underlying preference; Luce (1959) deepened the notion of probabilistic choice behaviour; and, Lancaster (1966) conceptualised alternatives as bundles of attributes. Below we discuss the various, often closely related, ways in which behavioural (choice) theory strengthens the current theory-driven choice modelling paradigm.

*3.1.1 Guidance in model specification*

Behavioural theory provides guidance to the analyst in the model specification phase. Discrete choice models typically employ handcrafted specifications where the analyst must determine how to specify the effect of each explanatory variable prior to model estimation. For example, should the effect of a particular variable on utility be linear or non-linear? Monotone or non-monotone? Is the effect moderated by other variables? The link to behavioural theories is valuable in that it offers the analyst a theoretical framework that can help guide and inform the process of how to include different explanatory variables within the model. For example, based on the theories on loss aversion, in situations where a reference point or situation is present, models are typically specified with a piecewise (linear) function which captures the difference between losses and gains (e.g. De Borger and Fosgerau 2008; Masiero and Hensher 2010).

In machine learning, guidance to the analyst in the model building phase does not come from (behavioural) theory. In fact, model building generally does not involve explicitly specifying the relationships that need to be estimated (relationships are 'discovered' as part of the process of training the model). Nonetheless, when building a machine learning model, the analyst still needs to make numerous decisions, regarding e.g., the performance function, activation function, hyper parameter setting, training algorithm, and the topology (e.g., the number of hidden nodes and connections in an ANN and the type of layers used). Model building therefore still involves a trial-and-error approach, in which competing models and set-ups are tested. Studies tend to be built on previous work. Therefore, benchmark data sets take a central role in model building in machine learning. Thus, while machine learning models are often pitched as generic models that can be used to model any problem, the 'best' model still needs to be discovered by the analyst by training and comparing numerous models.

Many machine learning models are actually partially knowledge based. In light of the adage *do not estimate what you already know* (Lennart 1999), much of the art of machine learning is determining how to incorporate problem specific knowledge into the model (Maren et al. 2014) – which is often a nontrivial task that requires substantial efforts by the analyst. Embedding problem specific knowledge is particularly important in case data and computational resources are limited. For instance, for ANNs the universal approximation theorem proves that an ANN with a single hidden layer consisting of a finite number of neurons can approximate arbitrarily well any real-valued continuous function (Hornik et al. 1989). But to achieve that in practice would require vast amounts of data and computational resources. By embedding problem specific knowledge, the network can more efficiently learn from the data, and thus achieve better performances given the available (limited) amount of data and computational resources. A good example of embedding problem specific knowledge is the use of so-called convolution layers in object detection tasks in computer vision applications. In convolution layers, the same weights are used when 'looking' at different parts of an image. This makes sense as –colloquially speaking– recognising an object (whether it is a dog or bicycle) in the upper left corner involves the same operations as recognising that object in the lower right corner. Using fully connected layers instead of convolution layers would be prohibitively expensive in terms of weights (note that typical computer vision models that make use of convolution layers already contain 10 to 100 million trainable weights).

*3.1.2 Meaningful model parameters for model validation*
Behavioural theory provides meaning to the model parameters (see also section 2.2). Parameters of a discrete choice model are not just 'meaningless' (regression) coefficients, but can be given a richer, behavioural, interpretation due to the link with behavioural theory (McFadden 1980; Hess et al. 2018). For instance, parameters of a Random Regret Minimisation (RRM) model can be interpreted as the *maximum* change in regret due to a unit change in the attribute level (the actual regret experienced by the decision maker depends on the performance of the considered alternative relative to other alternatives in the choice set) (Chorus 2012). Moreover, parameters of theory-driven discrete choice models are generally interpreted as estimates of population means. Accordingly, an estimated theory-driven choice model as a whole is understood to represent the decision making process of the average, or representative, decision maker. As this interpretation is conditional on the sample on which the choice model is estimated being representative for the target population (meaning it is a random draw from the target population), in the choice modelling field considerable attention is given to sample quality and ways to mitigate potential biases caused by less-than-ideal samples (Manski and Lerman 1977; Batley et al. 2017).

As the parameters in theory-driven discrete choice models carry behavioural meaning, they can also be used to validate an estimated model. That is, behavioural theory often provides expectations regarding the signs and relative magnitudes of parameters. For instance, in the context of a utility theory-based model, we expect the marginal utility of cost to be negative; and the ratios of the marginal utility of time and the marginal utility of cost – which reflects the value-of-travel time – to be close to the average wage rate (Small 2012). Similarly, in a gain-loss setting, based on loss aversion theories we expect the marginal (dis)utility associated with losses to be larger than the marginal utility associated with (equivalently sized) gains (Kahneman et al. 1991; Masiero and Hensher 2010). If the estimated relationships are consistent with theoretical expectations, then the model is validated – or at least a necessary step towards model validation is taken– and can be trusted to make inferences and predictions.

In machine learning, model parameters cannot be given a physical or behavioural interpretation as many machine learning models are not theoretically identifiable (Watanabe 2009; Ran and Hu 2017). In the absence of a theory to provide physical or behavioural interpretation to model parameters, the validity of the learned relationships cannot readily be established.[4] As a first validity check in supervised machine learning classifiers, the confusion matrix is usually inspected. But this does not offer any insight on the inner workings of the model. In fact, several recent studies have pointed out the drawbacks of limited model interpretability and explainability. For instance, Buolamwini (2019) shows that machine learning systems have learned systematic racial and gender biases, often picked up as reflection of underlying patterns of discrimination that exist in the real world. In response, model explainability has gained recent prominence in the machine learning field (Burkart and Huber 2021). For a wider adaptation of machine learning in domains where impacts of decisions can be critical, such as in health care settings, model explainability is a prerequisite. In the context of governmental decision making, model explainability is often not only desirable, but also legally required (Bibal et al. 2021).

In light of this, a wide range of so-called Explainable AI (XAI) techniques have recently been developed. XAI techniques enable the analyst to examine whether the model has learned intuitively reasonable relationships, as opposed to spurious, inexplicable or otherwise undesirable ones. Burkart and Huber (2021) distinguish five mechanisms for model explanation. Some techniques aim for global interpretation of the model, others for local (i.e. explanations for a specific prediction). Some techniques are model agnostic, meaning they can be used for any type of model, others are specific to a particular type of model. Examples of XAI techniques are Class Activation Maps (CAMs) (Zhou et al. 2016), Activation Maximisation (AM) techniques (Erhan et al. 2009), Layer-wise Relevance Propagation (LRP) techniques (Bach et al. 2015), Local Interpretable Model-agnostic Explanations (LIME) (Ribeiro et al. 2016), SHapley Additive exPlanations (SHAP) values (Lundberg and Lee 2017), and see Burkart and Huber 2021 for an overview).

---

[4] A notable exception is Probabilistic Graphical Models, which exploit parametric functional forms. Just like Structural Equation Models, PGMs embed a theoretical model, which is a priori imposed by the researcher. Accordingly, PGMs provide interpretable components in the form of posterior distributions.

However, XAI techniques are rarely straightforward to use and their application often requires significant additional effort on the part of the analyst. In light of this, some machine learning experts have argued that elucidation techniques may not be the best approach to validate and build trust in machine learning models, and that the way forward is to design models that are inherently interpretable (Rudin 2019). In the choice modelling context, so far only a few XAI techniques have been tried. For instance, Alwosheel et al. (2019) and (Alwosheel et al. 2021) have pioneered the use of, respectively, Activation Maximisation and Layer-wise Relevance Propagation techniques to assess the validity of the relationships learned by ANNs trained on travel model choice data. At present, it is however unclear to what extent XAI techniques are able mitigate the opaqueness of machine learning models for choice modelling.

*3.1.3 Economic outputs*
Behavioural theory enables the derivation of rigorous economic outputs. This is particularly true for discrete choice models grounded in Utility Theory (UT). In RUM models in linear-additive form, the ratios of parameters directly yield their marginal rates of substitution (Ben-Akiva and Lerman 1985). Arguably the most widely used marginal rate of substitution derived from theory-driven discrete choice models is the Value-of-Travel Time (VTT) (Small 2012). Moreover, due to seminal theoretical works by e.g., Small and Rosen (1981) and McFadden (1981), firm connections between UT based choice models and welfare theory have been established. These studies have shown that the choice probability function of RUM models can be considered as the expected uncompensated demand curve of a particular alternative; and that the change in consumer surplus –which is a pivotal notion in welfare theory– due to an intervention in the set of available alternatives in a choice situation can be computed using the so-called change in the logsum formula (cf. de Jong et al. 2007). This elegant connection enables a 'seamless' transfer of such theory-driven discrete choice model outputs to other economic models and appraisal methods grounded in utility theory, such as Cost Benefit Analysis (Mackie et al. 2014).

In contrast, there is no established framework to derive economic outputs from machine learning models. However, recognising the value of economic outputs provided by the connection with behavioural theory in discrete choice models, several scholars have recently attempted to tie machine learning classifiers with economic theory. For example, decision trees have been used to mimic non-compensatory behaviours, consistent with the elimination-by-aspects heuristic (e.g. Arentze and Timmermans 2004; Arentze and Timmermans 2007; Brathwaite et al. 2017). Hidden Markov models that have found widespread application for speech recognition are being used by choice modellers to capture the effects of habit and inertia (Goulias 1999; Choudhury et al. 2010; Xiong et al. 2015; Zarwi et al. 2017). And, ANNs, which have emerged as the workhorse model for pattern recognition tasks, are increasingly being blended with random utility frameworks to leverage the benefits of both frameworks (Van Cranenburgh and Kouwenhoven 2019; Sifringer et al. 2020; Wang et al. 2020b). But, none of these efforts have yet put machine learning methods on par with theory-driven choice models when it comes to economic outputs.

*3.1.4 Ability to forecast behaviour in new settings*

Behavioural theory provides discrete choice models a strong basis for forecasting in new settings. Forecasting in new settings, such as policy interventions, is the purpose of many applications of discrete choice models (Brathwaite 2018). For a model to be able to forecast in new settings, or more formally to generalise *out-of-distribution*, a model must be *causal* (Pearl 2009; Schölkopf et al. 2021). In theory-driven choice models, the causal structure between the variables and the dependent variable (i.e. the choice) is provided by the theory. The causal structure represents structural knowledge about the data generating process, thereby enabling out-of-distribution generalisation. Having said that, it is well-known that a model is as good as its underlying assumptions. The claim that theory-driven models are better suited for out-of-distribution prediction ultimately rests on whether the theoretical assumptions are reasonable in the context of the application (Reiss and Wolak 2007). A theory-driven model based on a poor theory will still do a poor job in terms of out-of-distribution generalisation. But, judging from the widespread use of discrete choice models based on utility theory for out-of-distribution forecasting, it seems reasonable to conclude that utility theory provides a strong foundation for forecasting in new settings. For instance, RUM models are used throughout the world in large-scale transport models (Daly and Sillaparcharn 2000). They are used to make (long-term) forecasts for the impacts of say a new road or railway connection on travel demand (Hensher and Button 2000; Van Cranenburgh and Chorus 2018). Although backcasting studies that could underpin how well these models have actually performed in forecasting under new settings are few and far between in the scientific literature, the general impression seems to be that these models have been helpful for transportation planners and decision-makers and have been relatively accurate over the last decades (de Jong et al. 2008; Parthasarathi and Levinson 2010).

In the absence of behavioural theory, or other sources of structural knowledge about the data generating process, machine learning is less well equipped for forecasting under new settings. The vast majority of machine learning concerns non-causal statistical models – which contain less information about the data generating process than causal models (Schölkopf et al. 2021). Unsurprisingly in this regard, machine learning models have particularly been successful in, and used in the context of, short-term forecasts and prediction tasks, where the forecasting or prediction task context is expected to be within the support of the data used for training. In other words, machine learning excels when the forecasting conditions are identical to the training conditions. For example, the success of e-commerce platforms such as Amazon can at least partially be attributed to the power of machine learning algorithms for personalization and recommendation (Smith and Linden 2017). In most cases, these platforms have used some variation of collaborative filtering, an automated recommender algorithm that makes predictions about a particular user by drawing parallels with the behaviours of other 'similar' users observed in similar decision- making contexts (Sarwar et al. 2001; Linden et al. 2003).

Though it is theoretically clear that non-causal machine learning models are less suited for forecasting under new settings, there is little evidence in the form of empirical studies or Monte Carlo analysis that supports such a conclusion. To the best of our knowledge there are, for instance, no studies conducted using machine learning for making forecasts for new road or rail alternatives, or for long time horizons (10+ yr.).[5] A recent study by Toqué et al. (2017) explores using machine learning models for intermediate time horizon predictions and finds encouraging results using Random Forests and Long-Short Term Memory neural networks to forecast public transport travel demand for the business district in Paris´ Metropolitan Area one year ahead. But, no major changes in behaviour (i.e. the data generating process) occurred over the period of study. So, this forecasting study could still be considered as an example of forecasting within the support of the data used for training. Furthermore, from this study it is unclear how these machine learning based forecasts compare with conventional theory-based model forecasts.

### 3.2 Connection to stated choice experiments

Theory-driven discrete choice models often make use of data collected in Stated Choice (SC) experiments (Louviere et al. 2000). In fact, discrete choice models and SC experiments are so closely connected that they are frequently perceived as a single method. SC experiments have a number of attractive characteristics in and of themselves (thus irrespective of whether their data are analysed using theory-driven discrete choice models, or using some other model) (Cherchi and Hensher 2015). Firstly, SC experiments can be designed to answer very specific research questions, such as what the distribution of the value-of-time is, or to what extent does a new metro line pull travellers from active modes. Moreover, due to their hypothetical nature, SC experiments are particularly suited to study situations that do not yet exist in real-life. Consider, for the sake of illustration, the case of autonomous vehicles (AVs). AV technology is currently being trialled globally. When ready, its implications for existing patterns of travel and land-use behaviour are expected to be profound: some have predicted the end of private car dependent Western societies, others have portended greater suburbanization than has ever been observed before (e.g. Firnkorn and Müller 2015). Despite their well-known limitations, such as hypothetical bias and strategic behaviour (c.f. Fifer et al. 2014), SC experiments offer an attractive way to learn about choice behaviour in such an AV future (e.g. Correia et al. 2019). Secondly, SC experiments are fast to set-up and run. Nowadays, conducting an SC experiment is greatly facilitated by dedicated software packages developed for creating experimental designs (e.g. Ngene and STATA), and by online survey platforms (e.g. Qualtrics and SurveyMonkey) for swift online implementation of the SC experiment. Furthermore, online panel companies, such as Kantar and Amazon MTurk, enable quick access to respondents belonging to the target population. Thirdly, the analyst typically has full access, control and ownership over the data. Therefore, with SC data there is relatively limited exposure to other parties (e.g. to those who must give access to the data). This makes collecting SC data a relatively hassle-free option. Additionally, as the data are typically of relatively modest size, handling the data does not require special technical skills. Finally, privacy concerns regarding storage and publication of the data are usually limited. The respondents who subscribe to a panel company are aware their data are used for a wide variety of reasons, including scientific publication, and care has been taken by the panel company that the data cannot be traced back to individuals.

---

[5] Relatedly, there are few applications of machine learning model used for forecasting *aggregate* travel demand (a notable exception is Kostic et al. (2021)).

Discrete choice models and SC experiments are a particularly good marriage for, at least, the following two reasons. Firstly, together they offer a lean and elegant approach to acquire in-depth insights on preferences and choice behaviour. SC experiments provide a clean experimental setting that neatly fits the stylised way in which choice behaviour is modelled in theory-driven discrete choice models. Secondly, together they offer a comparatively inexpensive way to acquire in-depth insights on preferences and choice behaviour. As choice models consume only a modest number of parameters, a carefully designed SC experiment requires only a modest number of respondents to participate in the SC experiment to obtain statistically significant model parameters. Moreover, the experimental design of the SC experiment can be generated such that they are optimised for efficient estimation of the model parameters of discrete choice models (Rose and Bliemer 2009). This further decreases the number of respondents needed, and hence lowers the cost of data collection.

In contrast, machine learning and SC experiments are a less natural fit. Firstly, machine learning models often require larger data sets, especially the nonparametric ones. For instance, Alwosheel et al. (2018) find that as a rule-of-thumb roughly 50 observations are needed for every weight in a shallow fully connected ANN. This means conducting SC experiments aimed to be used in combination with machine learning models instead of theory-driven choice models tend to require comparatively more respondents (although not to the extent that the combination becomes infeasible – see our literature overview in section 2.3 on the use of SC data). Secondly, and arguably more decisive, the clean experimental setting of SC experiments offers limited scope for machine learning models to outperform their comparatively less flexible theory-driven counterparts. SC experiments are designed to 'measure' choice behaviour under highly stylised hypothetical conditions controlled by the analyst. Therefore, unexpected context effects that are likely to arise in real-life settings e.g. stemming from interactions with numerous external factors, such as weather conditions or temporary road closures in the context of transport behaviours, are minimised or not present at all. Likewise, the use of decision heuristics –which in real-life may be triggered when a decision maker is confronted with a large choice set (e.g. when searching for a flight online) are circumvented in carefully designed SC experiments comprising of not more than a handful of alternatives. Furthermore, the number of explanatory variables in SC data is usually modest, implying a limited scope for high order interactions to be at work. Altogether, this suggests that SC data –in its current form– are not the type of data where machine learning models can be expected to outshine their theory-driven counterparts by a large margin. But, the way is which SC data are collected may evolve over time. For instance, choice experiments conducted in virtual reality settings may produce large amounts of data with very little to no effort from the respondent (Farooq et al. 2018). For analysing data from such novel SC experiments, machine learning could be particularly useful.

Machine learning models are a more natural fit to large (passively collected) data sets (Farooq et al. 2015). In the context of ongoing digitisation of societies, passively collected data will become more and more abundant in the years to come. These data typically comprise many variables (and hence have many more potential interactions between them) that could co-explain (choice) behaviour. As a result, the flexible nature of machine learning models is likely to bring about larger gains in the context of these types of data than when applied to data from SC experiments. Recent studies by e.g. Sun et al. (2018) and Lhéritier et al. (2019) that use large revealed preference datasets seem to support this notion. They report noteworthy improvements in model performance as well as model specification and estimation time using machine learning models, compared to theory-driven discrete choice models.

In conclusion, theory-driven choice models have an edge over present day machine learning models in terms of interpretability, economic outputs, and forecasting in new settings. As these are pillars of the field, machine learning models in their present forms are –in our view– unlikely to replace theory-driven model on the short-term. However, machine learning has many strengths of its own too. By embracing machine learning the choice modelling field can improve its practices. The next section explores opportunities for doing so.

# 4  Opportunities for embracing machine learning models, techniques and practices for choice modelling

After having (re)discovered the core merits of the prevailing theory-driven modelling paradigm in section 3, in this section we explore areas of opportunity for embracing machine learning for choice modelling. We focus on five areas. In these areas, studies have already started applying machine learning methods to improve choice modelling, and some progress has already been made. The identified set of areas is by no means meant to be exhaustive. Inevitably, there are other areas of opportunity that we have overlooked.

## 4.1  Model building

The search for model specifications is perhaps the most heavily researched area within choice modelling. In their classic book 'Discrete Choice Analysis: Theory and Application to Travel Demand', Ben-Akiva and Lerman (1985, chapter 7.2) describe the process of model building as follows "*It [model building] is a mixture of applications of formal behaviour theories and statistical methods with subjective judgments of the model builder*". In spite of the guidance provided by behavioural theories (see section 3.1.1), there is seldom a prior 'optimal' model specification. Therefore, the analyst usually estimates a series of model specifications, with the aim to find the 'most appropriate' model specification. The analyst starts this process based on prior belief about the underlying data generating process and revises the model assumptions along the way based on the statistical evidence provided by confronting the model with the empirical data. Due to the many different specifications that can possibly be tested, this practice can be labour-intensive, and any search for the most appropriate specification is necessarily ad hoc (Keane and Wasi 2013; Vij and Krueger 2017).

Machine learning can help surmount some of the challenges related to model building. In the three sub-sections that follow, we discuss a number of exemplar ways in which machine learning is used, or could be used, for building choice models. In these studies, building choice models has become a process of mixing formal behaviour theories and applying machine learning tools.

*4.1.1 Finding utility functions*

Machine learning models and techniques can assist in finding the optimal utility function. Even when faced with a fairly limited number of explanatory variables, the number of testable utility functions is very large considering the various ways in which variables can be treated, such as using log and Box-Cox transformations, dummy coding, piecewise linear representations, discretisation, etc. To resolve this problem, Rodrigues et al (2019) capitalise on techniques developed in machine learning for 'feature selection'. By leveraging the Bayesian framework and the concept of automatic relevance determination (ARD), they automatically determine an optimal utility function from an exponentially large set of possible utility functions in a purely data-driven manner. Their method, which they call DCM-ARD, receives a set of possible variables in the utility function, and returns the posterior distributions on the "relevance" parameters for each variable or interaction term. High positive values indicate that the term likely exists in the optimal utility function. This follows the tradition in machine learning on regularisation-based feature selection, particularly the Least Absolute Shrinkage and Selection Operator (LASSO) method (Hastie et al, 2015).

*4.1.2 Capturing systematic heterogeneity*

Machine learning models can also directly be applied to capture systematic heterogeneity, i.e. tastes that vary systematically with observable variables. In theory-driven discrete choice models, such systematic heterogeneity is typically captured through specifying *interactions* between different pairs of variables in the utility function, such as e.g. an interaction between income level and cost. However, the number of testable higher-order interactions grows exponentially with the number of explanatory variables in a dataset, and a manual comparison across different utility specifications can quickly prove infeasible. To overcome this challenge, several scholars have used ANNs (e.g. Han et al. 2020; Sifringer et al. 2020; and Wang et al. 2020a). These models aim to discover the most appropriate specification, including high-order interactions, from the data as part of the process of model training. Another approach is taken by (Martín-Baos et al. 2021), who propose using Kernel Logistic Regression (KLR) models –a nonparametric extension of linear logistic regression models– and reinterpret KLR models within the RUM framework. What all these approaches have in common is that they propose model structures in which one part of the model is restricted and behaviourally interpretable through the use of an MNL kernel (i.e. softmax output layer), while another part of the model is flexible (i.e. capable of capturing interactions and/or nonlinearities) but not interpretable. Hence, the model structures in these approaches modestly restrict and/or impose structure in such a way that theoretical or behavioural relationships are (partially) preserved, or incorporated.

*4.1.3 Capturing random heterogeneity*

Machine learning models can be used to capture random heterogeneity (i.e. the variation in tastes across respondents that is not associated with observable variables). Several theory-driven model types have been developed to capture random heterogeneity, using discrete or continuous mixing distributions, or a combination thereof. Latent Class (LC) models use a discrete mixing distribution as a way to identify relatively homogenous consumer segments that differ substantially from each other in terms of their preferences (Kamakura and Russell 1989). LC models are in theory able to mimic any distribution to any arbitrary degree of accuracy. However, a drawback of LC models is that the number of segments must be defined by the analyst prior to model estimation, and the appropriate number of segments for any sample population is typically determined based on post-hoc model comparisons. Mixed Logit (ML) models are another, and probably the most widespread, model type to capture random heterogeneity. In ML models a continuous parametric mixing distribution is assigned to one or multiple model parameters to capture the distribution in tastes across decision makers. However, a potential issue with ML models is that there are only so many parametric distributions available to the analyst to test (e.g. uniform, normal, lognormal, logistic, exponential), and these may not match the true underlying distribution well. This, in turn, could lead to erroneous inferences (Hess et al. 2005). To overcome these issues, some researchers have proposed semi-nonparametric mixing distributions within the RUM-ML modelling framework as a way to incorporate random taste heterogeneity, such as the mixture of normals distribution proposed by Fosgerau and Hess (2009) and Bujosa et al. (2010), and the polynomial series expansions employed by Fosgerau and Bierlaire (2007) and Bastin et al. (2010). However, they are still limited in that they require the analyst to specify the shape and complexity of the distribution prior to estimation, as defined, for example, by the number of mixture components in the case of the mixture of normals distribution, or the order of the polynomial in the series expansion.

In response to the above challenges related to capturing random heterogeneity, several studies have developed approaches inspired by machine learning methods. For instance, Sfeir et al, (2021) use Gaussian process models in a Latent Class (discrete mixture) modelling framework. They use the Gaussian process model to model the class-memberships, while the classes themselves are specified as conventional linear-in-parameters RUM-MNL models, thereby preserving the strong feats of RUM based discrete choice models. In a similar way, Ruseckaite et al. (2020) overcome the limitations of the above mentioned semi nonparametric approaches through the development of Gaussian process mixture models where the distribution of taste parameters is specified to be a smooth continuous function, like semi nonparametric mixing distributions, but the shape and complexity of the distribution is estimated endogenously by the model. Relating to the challenge to determine the number of segments in LC models, Burda et al. (2008), De Blasi et al. (2010), Li and Ansari (2014) and Krueger et al. (2018) have developed Dirichlet process mixture models of discrete choice –an infinite generalization of traditional LC choice models based on Bayesian nonparametric methods (e.g. Neal 2000)– where the appropriate number of classes is discovered endogenously by the model. Following a different approach, Van Cranenburgh and Kouwenhoven (2020) propose an ANN based method to capture random taste heterogeneity in the context of panel data obtained from two-alternative-two-attribute SC experiments, which capitalises on the behavioural notion of indifference.

## 4.2 Model evaluation and selection

In theory-driven choice modelling, evaluation and selection is not a clear-cut process with a set of algorithmic rules. More parsimonious models are generally preferred over less parsimonious ones. That is, in case two models perform equally well in the statistical sense, the principle of Occam's razor is applied, dictating that parsimonious models are preferred over more complex (i.e. less parsimonious) models. Statistical comparisons of goodness-of-fit are mostly based on in-sample performance, as determined by measures such as the Akaike Information Criterion (AIC) and Bayesian Information Criterion (BIC) (Kuha 2004). However, statistical performance is not the only criterion that matters in model evaluation and selection; congruence with behavioural expectations –as captured by the sizes, ratios, signs, and statistical significances of the model parameters– carries considerable weight as well. Thus, when evaluating models, a theory-driven choice modeller has to trade off goodness-of-fit for behavioural realism. For instance, a three-class latent class model can well be chosen over a better fitting five-class latent class model that is behaviourally less appealing (e.g. Mouter et al. 2017).

In machine learning, model evaluation and selection are comparatively more unidimensional. As behavioural realism or concurrence with theoretical expectations are not a criterion, the process of model evaluation and selection is almost exclusively based on statistical performance. Statistical performance is usually established through out-of-sample goodness-of-fit measures (Murphy 2012). That is, the model that obtains the best performance on out-of-sample data is said to have the greatest generalisability and is therefore considered to be the best model. Due to the often large number of parameters, a core commandment in machine learning is to split any dataset into a training set and a test set: *thou shalt not estimate your model with the same observations that you use to validate it*. There is multiple widely used out-of-sample evaluation techniques in machine learning, such as the use of hold-out samples and k-fold cross-validation (Murphy 2012). Moreover, the widespread use of benchmark datasets in machine learning, such as ImageNet (Deng et al. 2009) and MNIST, facilitates comparing model performance across studies and keeping track of progress made over the years in the field.

There is scope in choice modelling to improve its model evaluation practices by embracing techniques developed and popularised in machine learning, in particular those concerning the evaluation of statistical performance. The choice modelling field could use out-of-sample techniques for model evaluation and selection more often than it currently does, instead of fully relying on in-sample statistics as is prevailing practice. In their review of peer-reviewed behavioural studies in transportation published in the last five years, Parady et al. (2020) find that 92 per cent of the studies report some sort of in-sample goodness-of-fit statistic, but only 18 per cent report any equivalent out-of-sample validation measure. While these in-sample goodness-of-fit statistics and out-of-sample assessment frameworks are often asymptotically equivalent (Watanabe 2010), with small samples and high model complexity they can yield divergent results. In particular, it is found that out-of-sample validation methods place more emphasis on generalisation and rely less on assumptions of normality than in-sample goodness-of-fit statistics. It is therefore recommended that choice modellers too use out-of-sample model evaluation techniques for model selection.

### 4.3 Model estimation

Estimation of theory-driven discrete choice models is mostly based on standard maximum likelihood methods, estimated using some version of gradient descent-based optimization algorithms, such as the BFGS algorithm, where at each iteration the algorithm computes the exact gradient of the likelihood function using the full data. Consequently, computational times are determined by both the size of the dataset and the complexity of the likelihood function; as either increases, so do computational times. This practice becomes increasingly unwieldy when faced with ever larger data sets.

Machine learning has made significant advances in estimation (training) algorithms capable of working with large volumes of data and complex model specifications. These algorithms can be employed for estimation of choice models. For example, Lederrey et al. (2021) have recently adapted stochastic gradient descent methods for the estimation of discrete choice models, where at each iteration the algorithm computes an approximation to the gradient using a randomized subset of the full dataset. This approach is inspired by the mini-batch stochastic gradient descent methods used in machine learning, in which the parameters are updated after each batch instead of after processing the whole dataset (see Ruder (2016) for an exhaustive review of these and other gradient descent methods). Using mini-batches is found to reduce the variance and lead to more stable convergence. Furthermore, in machine learning GPUs are extensively used to perform matrix-based computations within optimization algorithms. In fact, the development of computing capabilities of GPUs has played a major role in recent advances in machine learning. Therefore, it seems worthwhile to explore the potential of GPUs for estimation of advanced choice models.

## 4.4 Raw unstructured data types and sources

Studies estimate that in 2017 the world produced 38 megabytes of data per person per day (Raj 2014). Much of these data come in raw unstructured forms, such as images, videos, text and speech. Moreover, occasionally these unstructured data are continuously generated, as opposed to being collected during a fixed time period. Unstructured means that the data are usually stored in its naïve format, are not easy to analyse (e.g. to sort or rank), and that the data entries do not correspond to variables having pre-defined relationships (Gandomi and Haider 2015). For instance, while a whole image has structure (to a human), a single pixel in an image does not provide any meaningful information regarding whether the image is a dog, a bicycle or a traffic light.

Though much of these unstructured data involve human choice behaviour, they are seldom analysed using theory-driven choice models. The reason for this is that theory-driven choice models are not capable of handling such unstructured and often high-dimensional data, and/or their continuous flows. Theory-driven choice models, as well as most other causal models, rely on hand-engineered data in which semantically meaningful high-level variables are constructed by the analyst prior to the modelling, such as product quality level, crowdedness level, comfort level, congestion level, noise level, etc.

In contrast, unstructured and high-dimensional data can be handled and analysed with various machine learning models, such as Natural Language Processing (NLP) for text data and Computer Vision (CV) models for images and video data. Machine learning models can be efficiently trained on both raw unstructured data and hand-engineered variables (features) (Schölkopf et al. 2021). Below, we discuss developments in our field and opportunities that we see for using ML models capable of handling raw unstructured data types and sources for choice modelling.

*4.4.1 Text data*

Several studies have used textual data in the analysis of choice behaviour. For instance, Glerum et al. (2014) use responses to semi-open questions, where the respondent had to provide adjectives for difference public transport and active modes – which were later converted to ratings and then incorporated in a hybrid choice model. However, to the best of the authors' knowledge no study has yet incorporated written text data into choice models through the use of full-fledged NLP models. We believe there is considerable potential for this line of research, as highlighted by various successful applications of text data in the analysis of travel behaviour. For instance, Collins et al. (2013) use text mining to assess the quality of local public transport networks; Gu et al. (2016) use text to identify traffic incidents in real-time; Hasan and Ukkusuri (2014) use text data to model individual activity patterns over time; and, Baburajan et al. (2018) employ text data to predict intention to use new mobility services.

*4.4.2 Image and video data*

Image and video data offer an additional source of data for explaining choice behaviour. In fact, in many choice situations, ranging from buying a pair of shoes to choosing a tourism destination, as a decision maker it is hard to do without visual information. Unsurprisingly, numerous SC studies therefore use images (and to a lesser extent videos) to better represent the choice situation in more realistic ways, e.g. in the context of different planning options (e.g. Meyerhoff 2013; Mariel et al. 2015), cycling facilities (Griswold et al. 2018) and landscapes (Shr et al. 2019). However, in the absence of choice models that can handle images and videos in their full detail, the use of these data in SC experiments has been used cautiously because of the risk that they can convey too much information that, if not controlled explicitly, might generate responses that cannot be associated with the attributes that are being assessed in the model (Cherchi and Hensher 2015). Unsurprisingly in this regard, Shr et al. (2019) find that a choice model estimated on data from a SC experiment with visual information is more noisy (i.e. contains more unexplained variance) than the same choice model estimated on data from a similar SC experiment, but without visual information.

Some scholars have tried to bridge computer vision and choice modelling. Antonini et al. (2006) use a combination of CV models and choice models to predict facial expressions. To do so, they take a sequential approach in which they extract so-called expression descriptive units using an active facial appearance model (a type of CV model), which they then feed into an MNL model to predict the facial expression. More recently, Van Cranenburgh (2020) explores sequential and joint approaches to combine (pre-trained) Convolution Neural Network models (a widely used type of CV model) and MNL discrete choice models to model trade-offs between tax increases and landscape aesthetic values. We believe there is scope to further explore incorporating images and videos in choice models by embracing CV models. This potential is also highlighted by several recent related studies. For instance, Rossetti et al. (2019) use CV models to quantify perceptions of urban landscapes in street view images. Thereby, they are able to map perceptions throughout the whole city. Haghani and Sarvi (2018) use image processing techniques to analyse evacuation behaviour in the case of emergencies.

*4.4.3 Continuous choice data (data streams)*

Unstructured data sources are occasionally dynamic, offering continuous data streams that can be used potentially to update choice models in real-time. This is a significant departure from traditional choice modelling practices, where datasets are collected once (with the exception of repeated cross-section and longitudinal panels, but their use is rare), and models are estimated using all data collected up to the point of model estimation. As new information and communication technologies enable the easy collection of live streams of behavioural data, the choice modelling field has an opportunity to develop discrete choice models that are adaptive in real-time to new sources of information. In machine learning, the subject of online learning has received considerable attention (e.g. Nguyen et al., 2017; Anderson, 2008), and the methods developed therein can benefit the development of similarly adaptive online choice models. An exemplar study is Danaf et al. (2019), which leverages these principles to develop a hierarchical Bayes choice model framework for estimating and updating user preferences in the context of app-based recommender systems. In particular, their model estimates three sets of preference parameters – those general to the population, specific to an individual, and specific to a decision context – such that "the individual-level parameters are updated in real-time as users make choices" (ibid.) and this new information is fed back to the model.

## 4.5 Open science practices

Machine learning is an exemplar within the broader open science movement. According to The United Nations Educational, Scientific and Cultural Organization (UNESCO) open science is the movement to make scientific research and data accessible to all. This includes practices such as publishing open scientific research, campaigning for open-access and generally making it easier to publish and communicate scientific knowledge.[6] For example, the Journal of Machine Learning Research, one of the preeminent journals in the discipline, provides free online access to all published papers. Numerous datasets for machine learning are available in the public domain, such as the University of California, Irvine Machine Learning Repository and the ImageNet project. Most estimation software is usually available through open-source packages, such as TensorFlow (Abadi et al. 2016) and Scikit-learn (Pedregosa et al. 2011).

In the choice modelling field, there is considerable room to improve open science practices, and to learn from the machine learning field. In the choice modelling field, preeminent journals are not open access, data sharing and open software practices are few and far between, and a broad and pervasive culture of open science is lacking. Institutional barriers may have prevented the choice modelling field (and other fields) from adoption of open science as a fundamental tenet of research more so than it has in machine learning. But irrespective of the existence of such barriers, it is important to recognise that an open science practice is not inherent to machine learning any more than it is to choice modelling. We believe the choice modelling field should more strongly push for open science practices. Open science can particularly enhance model building and validation practices in our field. For instance, having a number of so-called 'golden' datasets would enable choice modellers to more easily compare the performance of a new model with those that have come before. This would allow the field to better keep track on progress that is being made over time. However, it is less clear to identify what kinds of choice modelling datasets would be most appropriate. In choice modelling the outcome of interest is often behavioural insight, rather than prediction accuracy. Datasets (and models) are far more often collected with a specific research question in mind. Nonetheless, there are contexts where golden datasets could be created. A good example being value-of-travel-time studies, where data typically have a very similar structure: comprising two alternatives with two or three attributes (cost, time and sometimes reliability). And even if the development of golden datasets is not workable, it is easy to see that an open data culture would increase the number of datasets analysed in papers, and thereby reduce the chances of reporting results that are specific to one particular dataset and not generalizable to other datasets.

---

[6] For a recent review of the benefits and challenges to open science, the reader is referred to Allen and Mehler (2019).

## 5 Conclusion and the road ahead

This paper has taken a first step towards consolidating knowledge on the use and value of machine learning models, techniques and practices for choice modelling. Specifically, we have (i) clarified the differences and similarities between the machine learning and theory-driven modelling paradigms (sections 2.1 & 2.2), (ii) reviewed the emerging body of literature of studies using machine learning for choice modelling (section 2.3), and (iii) identified areas of opportunity for embracing machine learning (section 4). Thereby, we hope not only to convince choice modellers of the merits of further integration of machine learning in our field, but also to further facilitate it.

This section finalises this discussion paper by looking ahead. We aim to provide an informed discussion on what research questions must be answered to understand if and how machine learning could transform choice modelling practices. By raising these questions, we hope to direct and inspire future research. As a vehicle-of-though, we float the idea that we could be at the start of a paradigm shift in choice modelling, triggered by machine learning. This thought is not unsubstantiated as in our review of the literature we concluded that our field is likely to be at the early stage of the process of integration of machine learning in choice modelling, consolidation has not yet started, and there are few signs of saturation of ideas for bringing machine learning and choice modelling closer together.

Therefore, let's try to see the current developments in our field in the context of Kuhn's classic ideas on the structure of scientific revolutions (Kuhn 2012). In Kuhn's seminal work, he distinguishes five phases of scientific change that jointly make up a paradigm change cycle. In phase 1 there is not yet a dominant paradigm; in phase 2 a paradigm establishes and what he coins as 'normal science' takes place. Phase 3 is characterised by a persistent inability to account for anomalies of the dominant paradigm, and by attempts to resolve these anomalies *within* the context of the dominant paradigm. In phase 4 the paradigm shift takes place, in which core assumptions are re-examined and a new paradigm establishes itself. In the last phase of the cycle (phase 5) the new paradigm has overthrown the old one, and 'normal science' is re-established. Moreover, Kuhn argues that –given the social context in which science takes place– a new paradigm must satisfy two requirements, namely the new paradigm must (1) resolve outstanding problems related to the previous paradigm, and (2) preserve a relatively large part of the abilities and accrued knowledge of its predecessor.

In Kuhn's cycle of paradigm change, the current developments in our field seem to best fit phase 3. In the choice modelling context, anomalies are not unexplainable empirical observations such as in the classic example of paradigm shifts in which the orbits of celestial objects could not satisfactorily be described by the Ptolemaic celestial system (triggering the Copernican heliocentric paradigm shift). Rather, in our context anomalies can be understood as persistent limitations of the theory-driven choice modelling paradigm. One example of such a persistent limitation is the inability to describe and explain behaviour in the presence of visual stimuli (see section 4.4.2). Despite wide recognition that visual stimuli are indispensable for choice behaviour in numerous situations, including residential location choice, tourism destination choice, and partner choice – to name a few, no satisfactory theory-driven models have been developed capable of incorporating visual stimuli. Current attempts to resolve this problem are made within the dominant paradigm, but without avail. As a result, choice modellers often are advised to use images sparingly (Cherchi and Hensher 2015). Another good example of a persistent limitation concerns the restricted ability of theory-driven choice models to account for social interactions between decision makers. To paraphrase McFadden –one of the founding fathers of our field– on this limitation "*their omission* [social interactions] *makes choice models incomplete and misleading*" McFadden (2010). Because of the abundance of evidence that humans are social beings that heavily influence, and are influenced by, each other, choice modellers and economists alike have long sought to integrate social contexts into this framework (e.g. Manski 1993; Brock and Durlauf 2001; Maness et al. 2015) – hence attempting to resolve the problem within the existing paradigm. But, these efforts have not led to major breakthroughs. Letting go of individual rationality turns out to result in major equilibrium and endogeneity problems (McFadden 2010), which in turn undermine one of the core strengths of the theory-driven modelling paradigm: its ability to yield economic outputs.

The big question thus is whether or not machine learning (in some form) can resolve these and other outstanding problems in choice modelling, while preserving the key abilities and accrued knowledge of theory-driven choice models. In other words, is the machine learning paradigm strong enough to overthrow the four-decade dominance of RUM-based choice models? At present, that question is impossible to answer. There are major unknowns that will (jointly) determine how transformative machine learning will be to our field. Below we try to formulate these unknowns into five major research questions. Each question can be traced back to the requirements for new paradigms set by Kuhn, being: (1) can the new paradigm resolve outstanding problems, and (2) can the new paradigm preserve current abilities and accrued knowledge.

*To what extent can we develop (new) models that successfully integrate the two modelling paradigms?* In section 4.1 we have seen examples of studies that integrate the two modelling paradigms – creating so-called hybrid models. Integration is usually motivated by the promise to leverage the benefits of both frameworks. Hybrid models often attempt to retain key abilities of the current paradigm through a choice model part while overcoming some of the outstanding problems through a machine learning model part. Therefore, the development of hybrid models often boils down to incorporating theoretical relationships in otherwise fully data-driven machine learning models, or –more or less equivalently– incorporate flexibility of machine learning in theory-driven discrete choice models. As an example of the latter, Sfeir et al. (2021) incorporate machine learning flexibility in discrete mixture latent class choice models by using Gaussian process models for the class memberships allocation. Thereby, their model overcomes the rigidity of the mainstream logistic regression-based class membership allocation models, while maintaining behavioural interpretability and economic outputs provided by the linear-additive RUM-MNL kernels. As an example of the former, the model developed by Sifringer et al. 2020 splits an ANN into a fully flexible part and a highly constrained linear-additive RUM part. As a result, their model yields economic outputs related to the subset of the input variables that are processed by the RUM part. But the economic outputs are gained at the expense of the model being (slightly) less flexible than an otherwise fully connected ANN would have been.

As such, the development of new hybrid models will inevitably involve trade-offs between the pros and cons of both frameworks. It confronts our field with a new question, which is *how to strike the balance between the pros and cons when developing new hybrid models?* Future research needs to shed light on this question. Interestingly, the machine learning field is also increasingly interested in this question. In machine learning, the search for models that incorporate theoretical relationships is not new. For instance, as early as the 1990s, Joerding and Meador (1991) proposed two approaches to incorporate a priori information in ANNs based on architecture and weight constraints, which allowed imposing concavity and monotonicity (notably, both constraints were motivated from economic theory). However, these and other related ideas on incorporating theoretical relationships within machine learning have only recently gained traction (Rudin et al. 2021). Nowadays, an increasing number of machine learning researchers are convinced that the range of application of machine learning models that work purely based on associations is ultimately limited (Wager and Athey 2018; Rudin 2019). The incorporation of theoretical relationships can enable the development of causal models, and this is increasingly viewed as a desirable outcome, even if it comes at the expense of predictive performance.

*To what extent can Explainable AI techniques mitigate the opaqueness of machine learning models for choice behaviour modelling?* Most machine learning models are opaque, in the sense that it is often unclear – e.g. from looking at the learned model parameter (weights) – how they arrive at their predictions (see section 3.1). This opaqueness lessens machine learning models' immediate usefulness for choice modelling (and in turn their adoption) as in many choice modelling applications the aim is obtaining behavioural insights (thus inference). Recently, many new XAI techniques have been developed in machine learning (Burkart and Huber 2021) which potentially could take away, or at least lower, this obstacle to the use of machine learning models for choice modelling. However, while their value has been well established in machine learning (Molnar et al. 2020), the value of XAI techniques for choice modelling is still largely unclear. Only a few XAI techniques have as yet been pioneered in choice modelling (e.g. Alwosheel 2020). As a consequence, at present we have limited understanding about which XAI techniques are particularly promising and meaningful to our field, or to what extent they can mitigate the limitation of machine learning models' opaqueness for choice modelling. Relatedly, it is as yet unclear what sort of explainability (e.g. local or global explainability) is of importance for our field. Therefore, we do not know to what extent abilities of the current paradigm can be preserved in a machine

learning paradigm using XAI techniques. Future research needs to deepen our understanding of (the need for) explainability and XAI techniques' potential for choice modelling. For empirical work, the fact that many implementations of XAI techniques are available in open-source software repositories may be of help (Molnar et al. 2018; Nori et al. 2019; Anders et al. 2021)

*To what extent can we capitalise on the improved prediction performance of machine learning?*
Many studies motivate their use of machine learning for choice modelling based on the premise that machine learning can achieve higher prediction performance than its theory-driven counterpart. Many comparisons have been made, and most studies indeed find empirical support for this proposition (e.g. Lee et al. 2018). We also know that machine learning models particularly have an edge in prediction performance when the prediction condition are identical to the training condition (typically short-term forecasting). But, at present there are crucial unknowns relating to how to capitalise on this potential gain in prediction performance in choice modelling applications. First of all, we lack a systematic understanding of the contexts under which meaningful gains in prediction performance of machine learning models can be expected. For instance, it seems intuitive that higher gains in prediction performance can be expected on RP data than on SC data (see section 3.2 for a discussion) as SC data are collected in a highly controlled setting, leaving less room for e.g. omitted variables and interactions that machine learning models could potentially pick up (see section 3.2). Likewise, it seems intuitive that gains in prediction performance can particularly be expected for data set sets with many explanatory variables, as opposed to just a few (either RP or SC). But, at present a systematic understanding regarding these and other context factors is missing. Secondly, we lack a thorough understanding regarding the sort of *practical* applications where improvements in prediction performance brought by machine learning models matter and outweigh the cost incurred with using them. Despite the growing number of studies using machine learning models for modelling choice behaviour, so far, few machine learning models have made their way towards applications outside of academia. For instance, despite the strong focus in the current body of literature using machine learning on modelling mode choice behaviour (see section 2.3) and the evidence of improved prediction performance in this context, we are not aware of any large-scale transport model in the world that has yet replaced its theory-driven mode choice model by a machine learning one. Lack of (meaningful) explainability and/or the need (or wish) for a model with out-of-distribution forecasting capabilities might be factors hampering its application in that context (e.g. Bhatt et al. 2020). Hence, it remains an open question for which practical applications the potentially enhanced prediction performance of machine learning models gives it a decisive edge.

*To what extent can we capitalise on machine learning models' ability to handle unstructured data?*
In this paper we have discussed (section 4.4) that machine learning models' ability to handle unstructured data such as text, image and videos, is a feature which theory-driven models critically lack. By capitalising on this capability, choice modellers could, for instance, start modelling demand for restaurants or places based on Google Reviews, analyse the importance of moral viewpoints of politicians for their voting behaviour based on their Twitter feeds, and deepen understanding of residential location choice using street-level images of neighbourhoods. At present incorporating unstructured data in choice models is still in its infancy (e.g. Otsuka and Osogami 2016; Van Cranenburgh 2020). Further research needs to explore the value and feasibility of machine learning models' capabilities to handle unstructured data for choice modelling.

*Do we need the flexibility of popular machine learning models for analysing choice behaviour?* Human choice behaviour is widely considered to be highly complex (Luce, 2005). Moreover, many choices are made in, or surrounded by, complex socio-technical systems, increasing the level of complexity of the decision making process even further. This all suggests that using highly expressive machine learning models – such as deep ANNs which are known to particularly perform well for highly complex learning tasks (Goodfellow et al. 2016) – to model the choice behaviour would show great improvement in model performance (e.g. in terms of model fit or hit-rate). However, the empirical evidence so far shows these highly expressive machine learning models do not dramatically improve model fit as compared to their much more parsimonious theory-driven counterparts. Future research therefore must address this paradox and shed light on the need for highly expressive and flexible models in choice behaviour modelling. This seems especially relevant since at present the most popular models in machine learning, such as ANNs and Random Forests, are also the ones that are used most commonly for modelling choice behaviour (see Table 2). Their use is often motivated by their successes in applications in adjacent domains (e.g. image classification), rather than by their appropriateness for modelling choice behaviour.